\documentclass[
    ,final            
  ]
  {aipproc}

\layoutstyle{6x9}

\begin{document}

\title{Modeling the Dust Spectral Energy Distributions of Dwarf Galaxies}

\author{Suzanne C. Madden}{
  address={Service d'Astrophysique, CEA Saclay, Orme des Merisiers 91191 Gif-sur-Yvette France (smadden@cea.fr)}
}



\begin{abstract}
Recent efforts on the modeling of the infrared spectral energy distributions (SEDs) of dwarf galaxies are summarised here. The characterisation of the dust properties in these low metallicity environments is just unfolding, as a result of recently available mid-infrared to millimetre observations. From the limited cases we know to date, it appears that the hard radiation fields that are present in these star-bursting dwarf galaxies, as well as the rampent energetics of supernovae shocks and winds have modified the dust properties, in comparison with those in the Galaxy, or other gas and dust rich galaxies.  The sophistication of the SED models is limited by the availability of detailed data in the mid infrared and particularly in the submillimetre to millimetre regime, which will open up in the near future with space-based missions, such as Herschel.
\end{abstract}
\maketitle


\section{Introduction}

Dwarf-like galaxies are thought to be the most popular actors playing
the role of building blocks in the hierarchical galaxy formation
theater. We have in our local universe a vast cast of dwarf galaxies
which purport to be analogs of these cosmologically important
characters. They generously provide convenient laboratories to peer
into the details of the interplay between star formation and the
interstellar medium (ISM) under conditions of low metallicity, as low
as 1/50 solar metallicity. The effects of low metallicity on the
ongoing processes in galactic environments, while poorly understood,
can be far reaching in terms of star formation, supernova feedback,
dust composition and evolution, molecule formation, galaxy morphology
and in general, all heating and cooling processes.  Starbursting dwarf
galaxies typically harbour super starclusters (SSCs), which are
compact, dense sites of starburst activity, and alert us to the fact
that while they are low mass galaxies, over the years, they have
gained the reputation for impressive star formation activity, often
concentrated in very young clusters. Yet how much of this activity is
hidden from us due to dust in these systems? These are important issues to address for the sake of characterising star
formation in the early universe.

The metallicity of the ISM of a galaxy evolves as a function of the
enrichment of the metals, in part, from the donation of stellar winds
from evolved stars, and more prolifically, due to supernovae events
\citep{dwek98}. Dust is also subsequently altered and destroyed due to
the harsh effects from stellar radiation and winds and shocks due to
supernovae. The observed SEDs of
galaxies are their fossil footprints, which, if deciphered accurately,
should unveil important characteristics of the history of the events of the
galaxy.

As recent as five years ago, a review on the modeling of the SEDs of
dwarf galaxies would not have been a concept.  Efforts in this
direction have been hampered by the intrinsically faint infrared (IR)
luminosities of these sources, in spite of the fact that many of them
are currently undergoing significant local starburst activity. Compounding this effect, there has been the preconceived notion that metal-poor
galaxies do not contain much dust to be concerned about and that
optical photons arrive to us unobscured, thus revealing
all. Therefore, historically, their physical properties have been
well-studied at optical wavelengths. The folly of ignoring dust
effects is becoming more and more apparent as studies of the dust and
gas properties of dwarf galaxies have advanced.
  
Here, I summarise some of the efforts that have been directed
specifically toward the modeling of the SEDs of dwarf galaxies.  While many other sophisticated SED models that exist today explain
the observations of a variety of galaxies, and can also be applied to
dwarf galaxies, I will focus on the direct efforts made
toward the interpretation of dwarf galaxies and which go
beyond the single or multiple grey-body only assumptions.
 
\section{The ``DUSTY'' model applied to dwarf galaxies}

The DUSTY model of \citet{ivezic+97} takes into account absorption,
emission and scattering by dust using self-similarities. All output is
dimensionless and must be scaled back to the observed SED. This method
makes for a clever mathematical convenience in solving the radiative
transfer problem with many parameters. With DUSTY, the central star
cluster is specified as well as dust composition and dust radial
distribution. DUSTY then calculates the radial distribution of the dust
temperature. Caution must be taken when using this model, since no
stochastic heating processes are invoked for the dust. All of the dust
is assumed to be in thermal equilibrium with the radiation field.

\begin{figure}[htbp]
  \centering
  \includegraphics[width=0.6\linewidth]{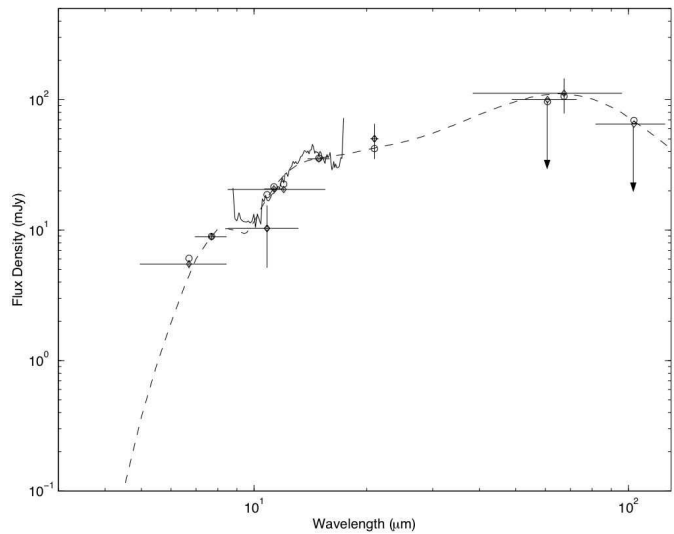}
  \caption{Results of the SED modeling of SBS0335-052 \citep{plante+02} using the DUSTY model \citep{ivezic+97}.}
  \label{fig:sbs}
\end{figure}
 
\begin{figure}[htbp]
  \centering
  \includegraphics[width=0.8\linewidth]{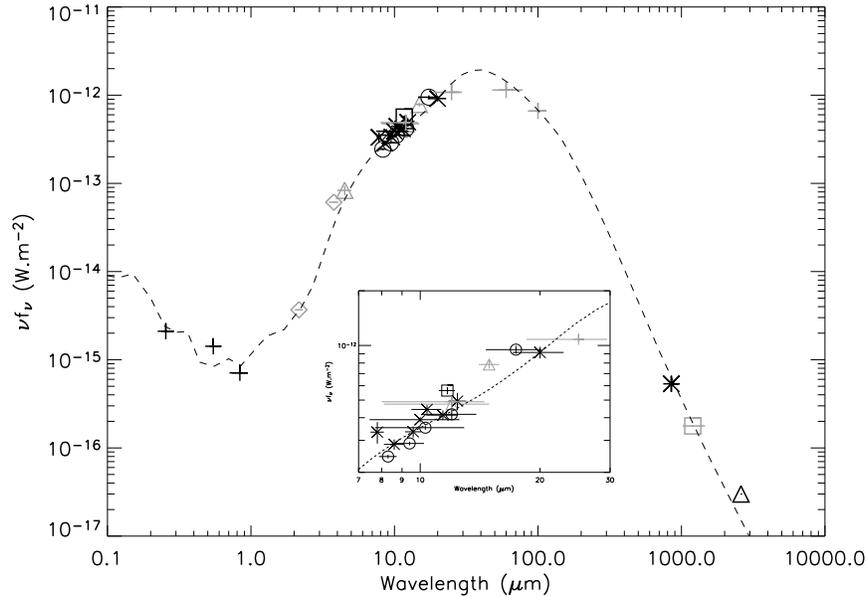}
  \caption{Results of the SED modeling of NGC5253 \citep{vanzi+04}, using the DUSTY model \citep{ivezic+97}. The inset contains a zoom into the MIR regime from 7 to 30$\mu$m.}
  \label{fig:ngc5253}
\end{figure}

The SEDs of SBS0335-052, (1/40 solar metallicity), and
NGC~5253 (1/6 solar metallicity) have been modeled using DUSTY
\citep{plante+02} \citep{vanzi+04}. SBS0335-052 was shown by ISOCAM
to harbor a significant amount of dust, due to absorption of the 9.7 $\mu$m silicate feature, accounting for $A_{V}$ = 30 to 40
\citep{thuan+99}. These mid infrared (MIR) results alone point to a
deeply-embedded SSC. Applying the DUSTY model to these two galaxies (Figure~\ref{fig:sbs}, Figure~\ref{fig:ngc5253}), the
authors note that the observations require dust with a flatter size
distribution, favoring larger grain sizes than those in our
Galaxy. The predominance of larger grains, and decrease of smaller
grains was explained as an effect of local, very young star
clusters. One possibility would be the fact that the high energy
densities in close proximity to massive young clusters may preclude
the survival of very small grains and PAHs, which also appear to be
absent. While silicate dust is obviously present in SBS0335-052, model
results point to the absence of silicate grains in NGC~5253, where
only carbon-based grains are used. The non-negligible dust masses that are determined
from the observed SEDs, attributed to a single cluster in each
case, are of the order of 10$^{5}$ M$_{\odot}$\footnote{Note that the dust mass in SBS0335-052 was overestimated in \citet{plante+02} because the 60$\mu$m ISOPHOT point was too high}. If
stochastic heating does prove to be an important process in these
environments, then the mass of dust will necessarily decrease. In both of these cases, the dust that is modeled using DUSTY is
perhaps characteristic of local environments around star clusters,
rather than dust that appears to be globally distributed throughout
the galaxy.

\section{A dust evolution SED model for dwarf galaxies}

Prompted by the possibility of dwarf galaxies being templates for
primeval galaxies, \citet{takeuchi+03} have created a dust SED model
which incorporates the dust evolution hypotheses of
\citet{hirashita+02}. The assumption for the SED model is that the
galaxies are very young ($10^{6}$ to $10^{8}$ yr) primeval galaxies. Thus
dust production is predominantly due to Type II supernovae. The dust
that is destroyed by supernova shocks is negligible, due to the youth
of the galaxy, compared to the dust formation rate. The far infrared
luminosity, L$_{FIR}$, and the dust temperature, T$_{dust}$, evolves
as a function of enrichment based on the supernova Type II grain
formation model of \citet{todini+01} and includes small stochastically-heated silicate
grains (a $< $10\AA) and bigger carbon grains (a $\sim$ 300\AA) in thermal
equilibrium with the radiation field.

\begin{figure}[htbp]
  \centering
  \includegraphics[width=0.6\linewidth]{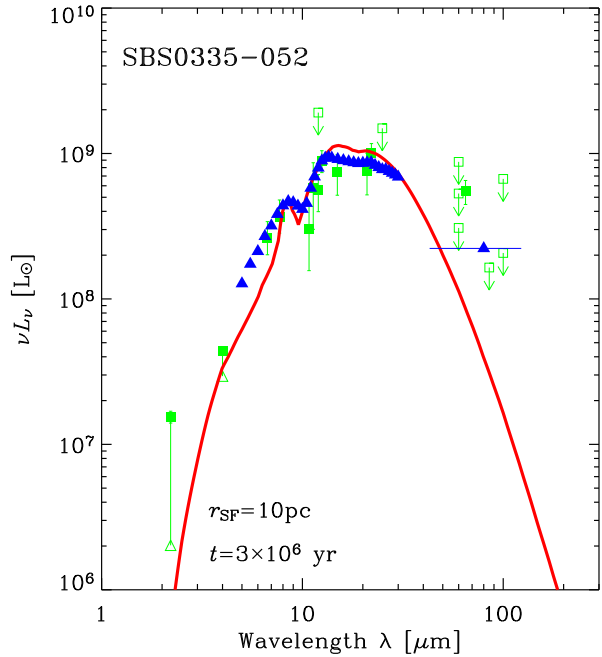}
  \caption{Results of the SED model of \citet{takeuchi+03} recently applied to SBS+0335-052 \citep{takeuchi+05a}. The squares are the observations cited in the original model of \citet{takeuchi+03} and the triangles are the new Spitzer data \citep {houck+04}. The solid line is the revised model fit to the observations.}
  \label{fig:takeuchi}
\end{figure}
 
\citet{takeuchi+03} applied this model to the extremely low metallicity galaxies SBS0335-052 (1/41 solar metallicity) and IZw18 (1/50 solar metallicity). The IR data for IZw18 is almost non-existent, but the model appeared to explain the available observations for SBS0335-052 well, concluding the presence of very hot dust (peak of 30 to 50 $\mu$m), in the first several Myr of its life. However, the original \citet{takeuchi+03} model (fitted to ISOPHOT data) deviates from recent Spitzer data longward of 20 $\mu$m. Now, the observed SED is flatter in the FIR, peaking toward shorter wavelengths ($\sim$ 30 $\mu$m) \citep {houck+04}. In order to account for the new Spitzer data, this model has since been improved by decreasing the radius of the effective star forming region, which correspondingly reduced the age of the starburst to $3x10^{6}$ years \citep{takeuchi+05a} (Figure~\ref{fig:takeuchi}). In the framework of \citet{takeuchi+03}, further improvements to the model have been made \citep{takeuchi+05b}, by modifying the size distribution according to the more sophisticated supernovae recipe of \citet{nozawa+03}, who take into account the radial density profile and temperature evolution of the the dust mass in the supernovae ejecta. The model can be applied successfully to primeval galaxies but has its limitations as the dust origins become complicated while the galaxy evolves in time. 

\section{Using the D\'esert et al dust model for dwarf galaxy SEDs}

The \citet{desert+90} dust model, originally used for the Milky Way
galaxy, assumes the dust emission and extinction is due to 3 different
dust components: the aromatic band carriers (PAHs), the very small
(1.2 to 13 nm) carbon grains (VSGs), and the classical big (15 to 110
nm) grains (BGs). Each component has a unique size distribution and
the model includes the process of stochastic heating. \citet{madden00}
first applied this model to explain the dust SED of dwarf galaxies,
allowing the grain size distributions to vary, depending on the
physical properties within the galaxies. This dust model was used, in
conjunction with a stellar evolution model (PEGASE: \citep{fioc+97})
and a photoionisation model (CLOUDY: \citep{ferland+98}) to develop a
self-consistent model to simultaneously explain the galactic
extinction and emission for the dwarf galaxies NGC~1569, NGC~1140,
IIZw40 and He~2-10 (\citep{galliano+03}; \citep{galliano+05}), with
metallicity values ranging from 1/6 to near solar metallicity.

\begin{figure}[htbp]
  \centering
  \begin{tabular}{cc}
    \includegraphics[width=0.5\linewidth]{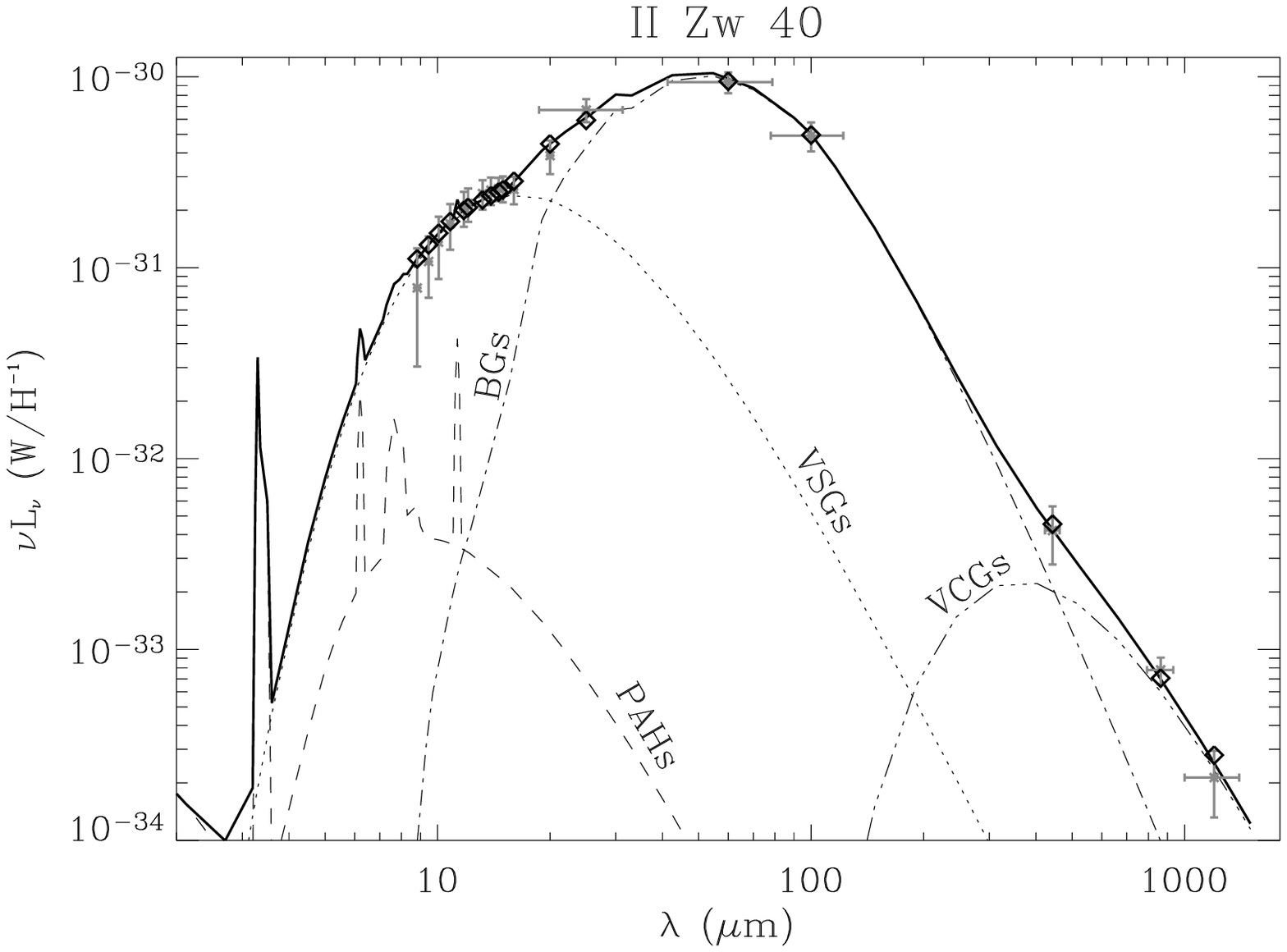} &
    \includegraphics[width=0.5\linewidth]{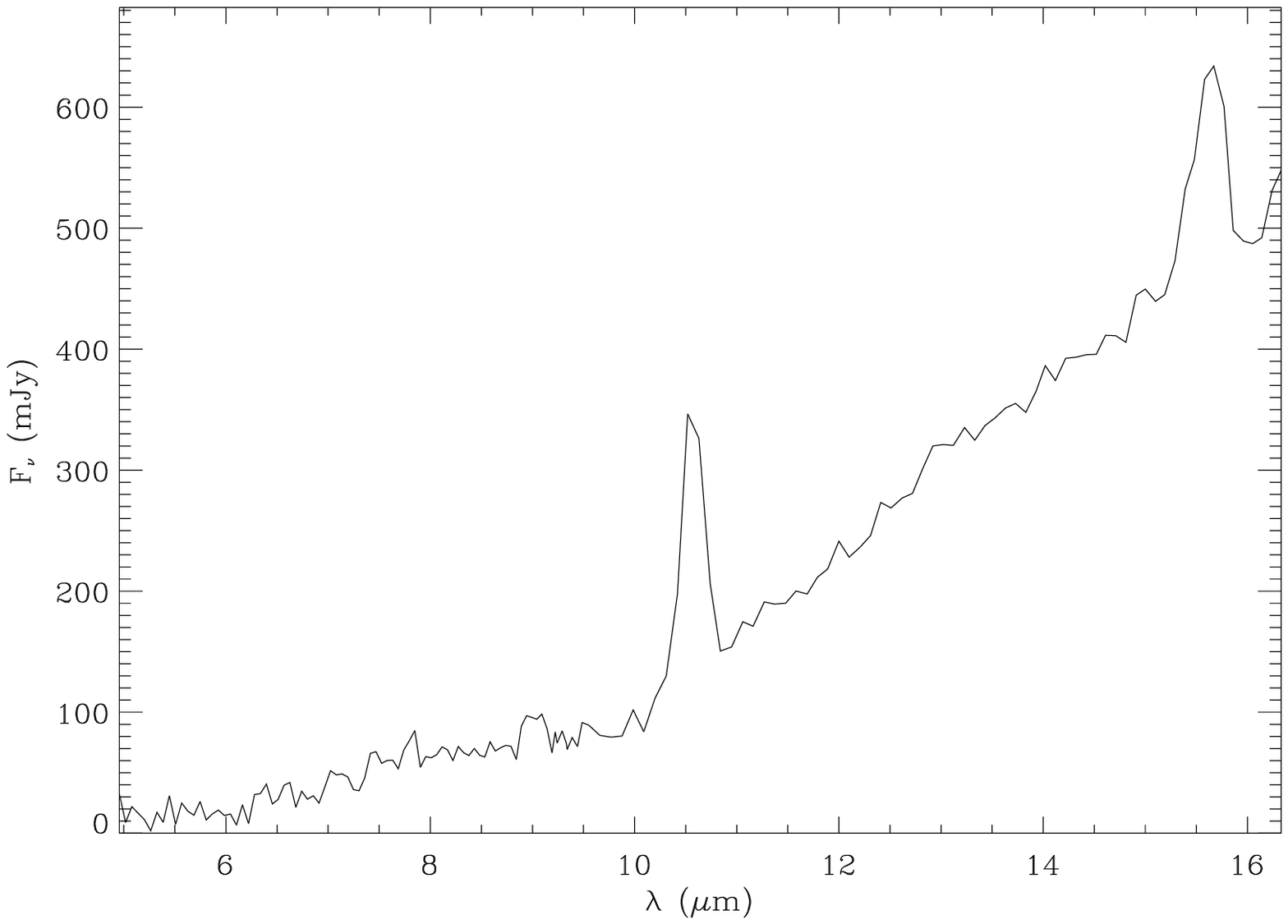} \\
  \end{tabular}
  \caption{Examples of the MIR to mm observed data and modeled SED (left: \citep{galliano+05}) and the MIR spectra for IIZw40 (right: \citep{madden+05}). Note the dearth of PAHs in the MIR spectra and the prominent [SIV]$\lambda 10.5\mu$m and the [NeIII]$\lambda 15.5\mu$m lines.}
  \label{fig:seds}
\end{figure}

The resulting dust properties of the modeled \citet{galliano+05} sample of dwarf galaxies are strikingly
different from those of the Galaxy and are dominated by small grains,
($\sim 3$nm) stochastically heated even at wavelengths as long as 100
$\mu$m in some cases (Figure~\ref{fig:seds}). This is in contrast to
results modeling local regions using the DUSTY code \citep{plante+02} \cite{vanzi+04}, where the dust size distributions
favored larger grains sizes, and a deficit of smaller
grains. Additionally, a cold dust component (T $\sim$ 7 to 10 K),
carrying at least 50\% of the mass of the galaxy, has been invoked to
explain the submm/mm excess observed in all of the galaxies modeled in such
detail \citep{galliano+03} \citep{galliano+05}. The presence of a cold dust component (< 10K) has also been observed in dwarf galaxies of the Virgo Cluster,
from ISOPHOT observations \citep {popescu+02}.
 
\citet{lisenfeld+02} also applied the \citet{desert+90} dust
model to interpret the observed SED of NGC~1569. Their approach
differs from that of the \citet{galliano+03} model, in that the dust size distributions for each component of NGC~1569 are fixed to those of the
Galaxy. Additionally the submm data sets are different and
not completely consistent, due to different calibration
procedures. The model of \citet{lisenfeld+02} assumes the spectral
shape of the Galactic radiation field, which is softer than the global
intrinsic radiation field as solved for by \citet{galliano+03}. While
their final fit shows an enhancement of the Galactic very small
grain component, relative to that of the big grain component, the
differences in the modeling approaches are sufficient to make it
difficult to properly compare the two models and results.

\subsection{SED Model Constraints: MIR gas and dust and dearth of PAHs}

One of the unique, important aspects of the detailed modeling
procedure by \citet{galliano+03} is pointing out the necessity of the
detailed MIR spectra in constraining the full dust SED as well as
offering the very important gas and dust spectral information in a
wavelength regime suffering little extinction. Diagnostic fine
structure nebular lines (e.g. [ArIII] 9$\mu$m, [SIV] 10.5 $\mu$m,
[NeII] 12.7$\mu$m, [NeIII] 15.6 $\mu$m ) as well as the 9.7 and 18
$\mu$m silicate bands and the PAH bands lie in the MIR regime.
\begin{figure}[htbp] 
  \includegraphics[width=0.7\linewidth]{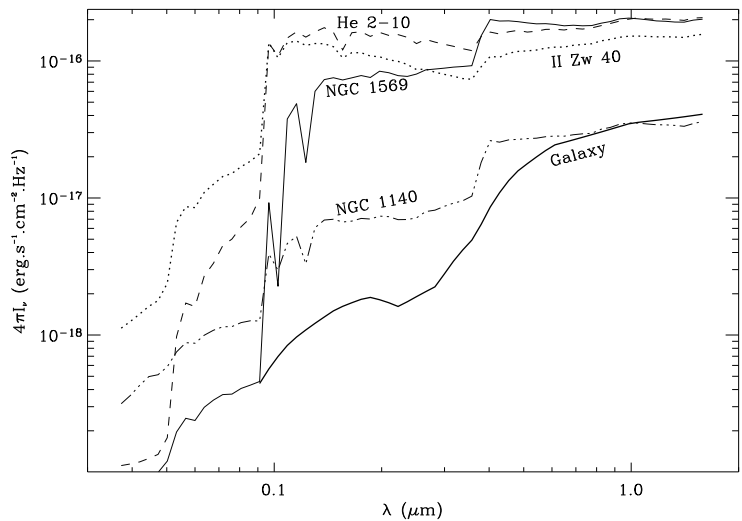} \\
 \caption{Examples of the variations of the
  modeled ISRFs for 4 of the dwarf galaxies, compared to the softer
  ISRF of the Galaxy.}
  \label{fig:isrf}
\end{figure}  

\begin{figure*}
  \includegraphics[width=0.9\linewidth]{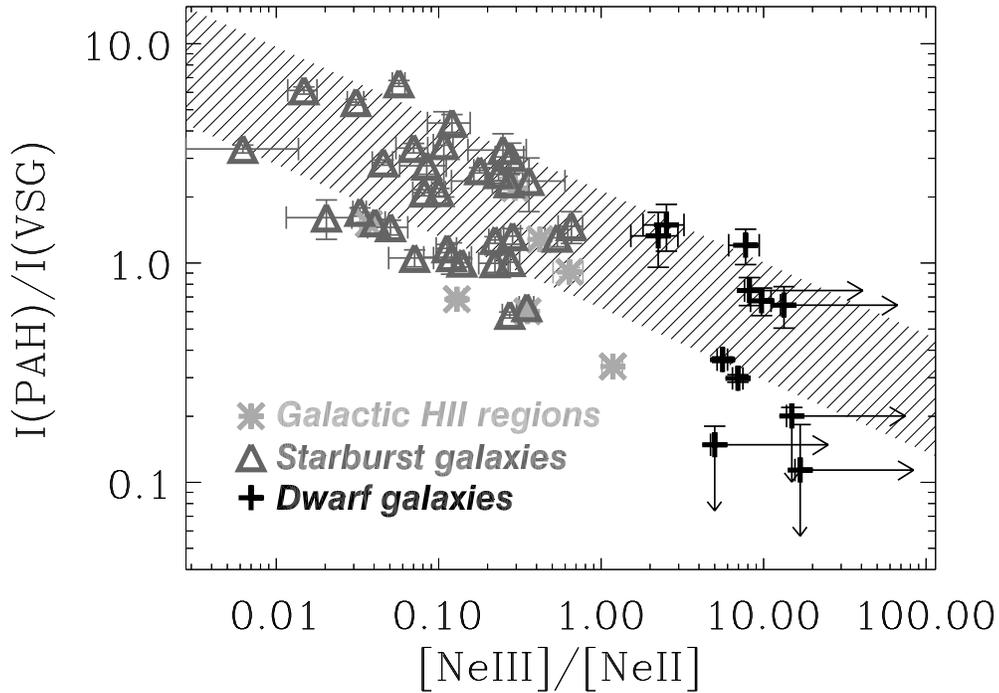} \\
 \caption{The correlation between the hardness of
  the global ISRFs indicated by the [NeIII]/[NeII] MIR line ratios and
  the PAH/15$\mu$m small dust continuum for a sample of HII regions \citep{peeters+02},
  metal rich starburst galaxies, and metal poor
  galaxies \citep{madden+05}.  As the ratio of the [NeIII]/[NeII] MIR
  lines becomes larger (e.g.$>$ 7 for dwarf galaxies), the PAH
  intensity drops remarkably, relative to the MIR continuum. }
  \label{fig:pahs}
\end{figure*}

As was initially shown in \citet{madden00}, PAHs are not a very
abundant dust component, globally, in dwarf galaxies (Figure~\ref{fig:seds}), in contrast to the global ISM of dust-rich starburst galaxies and spiral galaxies. Recent Spitzer data also confirm this \citep{houck+04}. To investigate the
absence of PAHs in dwarf galaxies, the hardness of the intrinsic modeled interstellar radiation
fields (ISRFs) of the sample from \citet{galliano+05} were
compared (Figure~\ref{fig:isrf}). As the
ISRFs become harder, the absence of PAHs becomes more
evident in the MIR spectra. Likewise, as the ratio of the [NeIII]/[NeII] MIR 
lines becomes larger, due to increasingly harder radiation fields (e.g.\citep{martin-hernandez+02}), the PAH intensity drops remarkably
relative to the MIR continuum (Figure~\ref{fig:pahs} \citep{madden+05}). For the dwarf galaxies, the values of [NeIII]/[NeII] $>$ 7, as compared to an order of magnitude less for Galactic HII regions. The hardness of the global radiation
field seems to be a factor that plays an important role in the
destruction of the PAHs \citep{madden+05}.  There is no obvious
correlation with metallicity and the absence of PAHs for the few
moderate metallicity cases here. However, the decrease of metals does result in a
larger mean free path length for photons. Thus on a global scale, the
hard photons could traverse larger areas, impacting larger scales in
these galaxies, subsequently destroying a larger volume of PAHs in the
galaxies \citep {madden00}.


\section{Summary of SED model results}

From the limited number of dwarf galaxy SED models that have been
constructed to date, each has its different origin, philosophy and
limitations. All of the results point to the presence of
non-negligible amounts of dust in low metallicity systems. The DUSTY
models applied to localised starforming regions in SBS0335-052 and
NGC~5253 (\citep{plante+02};
\citep{vanzi+04}), while not incorporating the process of stochastic
heating, conclude the tendency for larger grains to survive, rather
than smaller grains. In contrast, from global scale modelling
of a small sample of dwarf galaxies, using the dust model of
\citet{desert+90}, most of the grain populations are characterised by small grain sizes ($\sim$3 -4
nm) and are stochastically heated (\citep{lisenfeld+02}; \citep{galliano+03}; \citep{galliano+05}). Additionally, a large,
massive cold (T$\sim$5 to 10 K) dust component is an explanation for
the excess submm/mm emission observed the dwarf galaxy sample of 
\citep{galliano+03} and observed in other dwarf galaxies of the Virgo Cluster \citep{popescu+02}. The physical details of the dust as well as the quantity of dust present in these galaxies are important to understand in order to determine their extinction properties. Limitations in the
sophistication of the SED modeling efforts are related to the difficulty in
sampling the submm to mm wavelength regime with sufficient
detail. This will be remedied soon, with the launch of Herschel in
2007. Thus much further progress in modeling the SEDs of intrinsically
faint sources, such as dwarf galaxies, can be made.

\begin{theacknowledgments}
A special thanks to the organisers, Cristina Popescu and Richard Tuffs, for their hospitality and for a very stimulating, well-organised meeting.  The success of the work presented here is largely due to my collaborator, Fr\'ed\'eric Galliano. I thank Tsutomu Takeuchi for the more current figure of SBS0335-052, and for bringing me up to date on his new SED models. 
\end{theacknowledgments}


\bibliographystyle{aipproc}   

\bibliography{article2}
\IfFileExists{\jobname.bbl}{}
 {\typeout{}
  \typeout{******************************************}
  \typeout{** Please run "bibtex \jobname" to optain}
  \typeout{** the bibliography and then re-run LaTeX}
  \typeout{** twice to fix the references!}
  \typeout{******************************************}
  \typeout{}
 }

\end{document}